\documentclass[aps, preprint, superscriptaddress, prx, noeprint]{revtex4-2}

\usepackage{graphicx}
\usepackage{dcolumn} 
\usepackage{bm}
\usepackage{amssymb}
\usepackage{amsmath}
\usepackage{color}

\usepackage{xr}
\newcommand{\FigCap}[1]{\uppercase{#1}}	

\begin{document}

\title{Majorana-fermion origin of the planar thermal Hall effect in the Kitaev magnet $\alpha$-RuCl$_3$}

\author{K.~Imamura}
	\thanks{These authors contributed equally to this work.}
	\affiliation{Department of Advanced Materials Science, University of Tokyo, Kashiwa, Chiba 277-8561, Japan}
\author{S.~Suetsugu}
	\thanks{These authors contributed equally to this work.}
	\affiliation{Department of Physics, Kyoto University, Kyoto 606-8502, Japan}
\author{Y.~Mizukami}
	\affiliation{Department of Advanced Materials Science, University of Tokyo, Kashiwa, Chiba 277-8561, Japan}
	\affiliation{Department of Physics, Tohoku University, Sendai 980-8578, Japan}
\author{Y.~Yoshida}
	\affiliation{Department of Advanced Materials Science, University of Tokyo, Kashiwa, Chiba 277-8561, Japan}
\author{K.~Hashimoto}
	\affiliation{Department of Advanced Materials Science, University of Tokyo, Kashiwa, Chiba 277-8561, Japan}
\author{K.~Ohtsuka}
	\affiliation{Department of Physics, Kyoto University, Kyoto 606-8502, Japan}
\author{Y.~Kasahara}
	\affiliation{Department of Physics, Kyoto University, Kyoto 606-8502, Japan}
\author{N.~Kurita}
	\affiliation{Department of Physics, Tokyo Institute of Technology, Meguro, Tokyo 152-8551, Japan}
\author{H.~Tanaka}
	\affiliation{Innovator and Inventor Development Platform, Tokyo Institute of Technology, Yokohama 226-8502, Japan}
\author{P.~Noh}
	\affiliation{Department of Physics, Korea Advanced Institute of Science and Technology (KAIST), Daejeon 34141, Korea}
\author{J.~Nasu}
	\affiliation{Department of Physics, Tohoku University, Sendai 980-8578, Japan}
\author{E.-G.~Moon}
	\affiliation{Department of Physics, Korea Advanced Institute of Science and Technology (KAIST), Daejeon 34141, Korea}
\author{Y.~Matsuda}
	\affiliation{Department of Physics, Kyoto University, Kyoto 606-8502, Japan}
\author{T.~Shibauchi}
    \email{shibauchi@k.u-tokyo.ac.jp}
    \affiliation{Department of Advanced Materials Science, University of Tokyo, Kashiwa, Chiba 277-8561, Japan}

\date{\today}

\begin{abstract}
	{\bf The field-induced quantum disordered state of layered honeycomb magnet $\bm{\alpha}$-RuCl$\bm{_3}$ is a prime candidate for Kitaev spin liquids hosting Majorana fermions and non-Abelian anyons. Recent observations of anomalous planar thermal Hall effect demonstrate a topological edge mode, but whether it originates from Majorana fermions or bosonic magnons remains controversial. Here we distinguish these origins from low-temperature measurements of high-resolution specific heat and thermal Hall conductivity with rotating in-plane fields. In the honeycomb bond direction, a distinct closure of the low-energy bulk gap is observed concomitantly with the sign reversal of the Hall effect. General discussions of topological bands show that this is the hallmark of an angle-rotation-induced topological transition of fermions, providing conclusive evidence for the Majorana-fermion origin of the thermal Hall effect in $\bm{\alpha}$-RuCl$\bm{_3}$.} 
	\\
	\\
\end{abstract}

\maketitle

\newpage

Majorana particles are fermions that are their own antiparticles. The realization of Majorana fermions as quasiparticles, which describe the excitations from the ground state in quantum materials, is at the forefront of condensed matter physics. In particular, enormous efforts have been devoted to the search for the Majorana bound states using superconductors and their topological hybrid systems~\cite{Sato2016}, which are thought to be the first step towards the topological quantum computing based on Majorana-based non-Abelian anyons. 
A totally different approach to the Majorana physics is the exactly solvable Kitaev model for insulating honeycomb magnets with bond-dependent Ising-type magnetic interactions, in which the ground state is a Kitaev spin liquid (KSL)~\cite{Kitaev2006}. The excitations from the KSL ground state can be described by the itinerant Majorana fermions and $Z_2$ fluxes (or visons), which can be viewed as a fractionalization of the spins~\cite{Motome2020}. 

\begin{figure}[t]
        \centering
        \includegraphics[width=\linewidth]{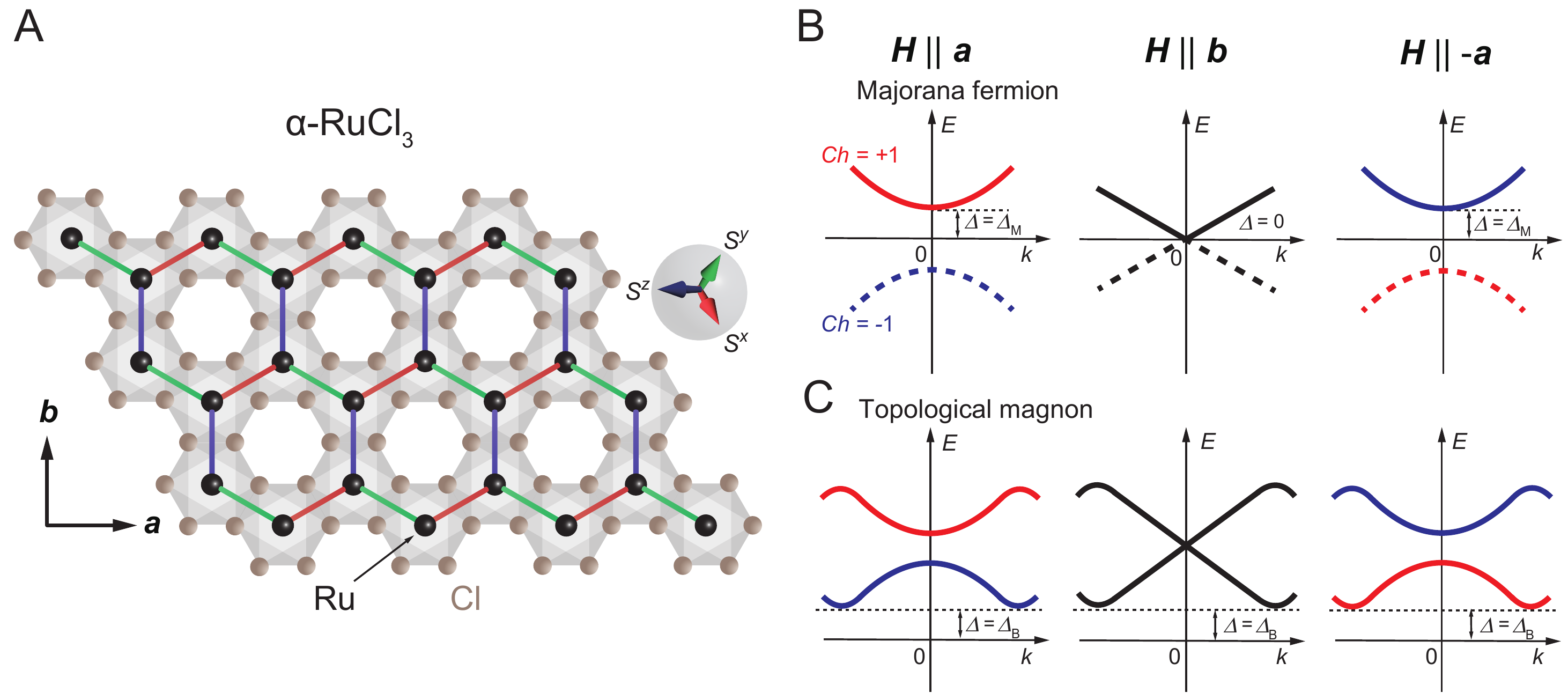}
        \caption{\label{fig:schematic}
		{\bf Key difference in the excitation spectra between Majorana fermions and topological magnons.} 
		({\bf \FigCap{a}}) Crystal structure of the honeycomb layer in $\alpha$-RuCl$_3$. Ru$^{3+}$ ions (black circles) surrounded by octahedrons (shades) of Cl$^-$ ions (brown circles) form a layered honeycomb lattice. We define $\bm{a}$ and $\bm{b}$ axes perpendicular and parallel to the Ru-Ru bond direction (zigzag and armchair axes), respectively (left inset). (Here we use the room-temperature $C2/m$ notation of the crystal axes within the honeycomb plane~\cite{bruin2022origin}.) 
		The Ising spin axes $S^x$, $S^y$, and $S^z$ (right inset) are, respectively, perpendicular to the plane including the red, green, and blue Ru-Ru bond and the crossing shared edge of Cl octahedrons. ({\bf \FigCap{b}}) Schematic energy dispersion $E(k)$ of the itinerant Majorana fermions in a KSL. The red and blue curves have different topological Chern numbers $+1$ and $-1$, respectively. Note that the $E<0$ part for the Majorana quasiparticle spectra is redundant (dashed lines), but its Chern number determines the topological properties. For field $\bm{H}$ parallel to the $\bm{a}$ (left) or $-\bm{a}$ axis (right), the Majorana excitations have a finite energy gap described by $\Delta_{\rm M}$, which closes for $\bm{H}\parallel\bm{b}$ (middle). In other words, when the field angle is rotated, topological band crossing occurs at {\em zero} energy in the $\bm{b}$-axis direction. ({\bf \FigCap{c}}) Similar plots for topological magnons. The band crossing occurs at a {\em finite} energy. In the spin disordered phase, the lowest energy band, whose Chern number determines the topological properties, always have a finite gap $\Delta_{\rm B}$. 
		}
    \end{figure}

The Kitaev interactions, in which three bonds connected to a honeycomb lattice point have orthogonal Ising spin directions, have exchange frustration, leading to the KSL ground state. Such frustrated interactions can appear through the Jackeli-Khaliullin mechanism~\cite{Jackeli2009} in real materials having edge-sharing octahedra surrounding magnetic ions. One of the prime candidates for these Kitaev magnets is a spin-orbit assisted Mott insulator $\alpha$-RuCl$_3$ with layered honeycomb structure (Fig.\,\ref{fig:schematic}\FigCap{a}). Each Ru$^{3+}$ ion with effective spin-$1/2$ has three bonds, where a cancellation of interactions between the two shortest Ru-Cl-Ru 90$^\circ$ paths leads to Ising interactions with the spin axis perpendicular to the plane including these two paths. Indeed, the existence of significant ferromagnetic Kitaev interactions with the energy scale of $\sim 5$\,meV has been reported by experiments in $\alpha$-RuCl$_3$~\cite{suzuki2021proximate}. Unfortunately, however, non-Kitaev magnetic interactions such as Heisenberg and off-diagonal terms also exist, and at low temperatures below $\sim 7$\,K an antiferromagnetic order with zigzag spin structure occurs in the absence of magnetic field~\cite{Johnson2015}. The application of in-plane magnetic fields of $\gtrsim 8$\,T can suppress the zigzag order, leading to the transition to a field-induced quantum disordered state with no magnetic ordering down to the lowest temperature~\cite{Yadav2016,Banerjee2018}. 

Within this high-field spin disordered state of $\alpha$-RuCl$_3$, the half-integer quantization plateau of thermal Hall conductivity $\kappa_{xy}$, which corresponds to a half of the quantum thermal conductance $K_0 = (\pi^2/3)(k_{\rm B}^2/h)T$ for electrons, has been reported~\cite{Kasahara2018,Yamashita2020,bruin2022robustness}. The quantized Hall effect is a signature characteristic to the edge current of fermionic particles, and the quantization to a half-integer value $K_0/2$ implies the non-Abelian phase of matter with Majorana fermions having a half degrees of freedom of electrons. Another remarkable feature in $\alpha$-RuCl$_3$ is that a finite transverse temperature gradient 
is observed even when the magnetic field is applied parallel to the in-plane thermal current, namely the emergence of anomalous planar thermal Hall effect~\cite{yokoi2021half}. In addition, the sign of quantized thermal Hall conductivity changes when the in-plane field component is reversed without changing the out-of-plane component. These results can be explained by the topological properties of Majorana fermions characterized by the Chern number $\pm1$, whose sign is directly related to the direction of the edge current and thus the sign of quantized $\kappa_{xy}$. The Chern number is determined by the field direction in the KSL~\cite{Kitaev2006}, which is consistent with the thermal Hall conductivity results~\cite{yokoi2021half}. 
However, it has been reported that the half-integer quantization has been seen in a limited temperature and field range which has slight sample dependence~\cite{Yamashita2020,bruin2022robustness,kasahara2022}. Moreover, in some samples the magnitude of thermal Hall conductivity $\kappa_{xy}$ does not reach $K_0/2$~\cite{czajka2021,czajka2023,lefranccois2022}. 

Recent theoretical calculations show that the sign change of the thermal Hall effect can also be explained by topological magnons in a specific parameter range, while the quantization is not expected~\cite{McClarty2018,Joshi2018,Chern2021,Zhang2021}. Very recent measurements of thermal Hall conductivity have reported unquantized $\kappa_{xy}$ in the high-field disordered state of $\alpha$-RuCl$_3$, and it has been claimed that the behavior of $\kappa_{xy}(T)$ could be explained by the topological bosonic mode of magnons~\cite{czajka2023}. It has also been suggested that a phonon Hall effect may be present and affect $\kappa_{xy}(T)$ in $\alpha$-RuCl$_3$~\cite{lefranccois2022}. Thus, at present stage, the variations in the $\kappa_{xy}$ data make it difficult to distinguish the origins, Majorana fermions and bosons, only from the thermal Hall measurements in $\alpha$-RuCl$_3$. In the bosonic scenario, $\kappa_{xy}\approx K_0/2$ would be just a coincidence. In the Majorana scenario, there are a few effects that can cause deviations of $\kappa_{xy}$ from $K_0/2$, including the phonon decoupling effect that can explain the temperature dependence~\cite{Vinkler2018,Ye2018} and the effect of partial cancellation due to twin domain formation that can account for the sample dependence of $\kappa_{xy}$ values~\cite{Kurumaji2023}. Under these circumstances, we need a different approach to distinguish these origins without relying on the magnitude of $\kappa_{xy}$. 

Here we point out that low-energy excitation spectra of Majorana fermions in the KSL and of topological magnons in the field-induced quantum disordered phase have essentially different field-angle dependence. This distinction between Majorana and magnon excitations is associated with the generic statistical difference that the Fermi energy is defined only for fermions and there is no negative-energy modes for bosons with zero chemical potential. In the Kitaev model~\cite{Kitaev2006}, the Majorana excitation gap $\Delta_{\rm M}\propto|M(\bm{H})|$ can be described by the mass function $M(\bm{H})=\frac{h_xh_yh_z}{\Delta_{\rm flux}^2}$, which has characteristic dependence on magnetic field $\bm{H}=(h_x,h_y,h_z)$ in the Ising spin axis coordinate. Here $\Delta_{\rm flux}$ is the $Z_2$ flux gap whose energy scale should be larger than the Majorana gap $\Delta_{\rm M}$~\cite{Motome2020}. We note that although the non-Kitaev interactions are present in $\alpha$-RuCl$_3$, these interactions give $h_x+h_y+h_z$ terms in the mass function, which become zero for the fields within the honeycomb plane~\cite{Tanaka2022,Hwang2022}. As the spin axes are different from the crystallographic axes (Fig.\,\ref{fig:schematic}\FigCap{a}), we expect $\Delta_{\rm M}\propto |\sin 3\phi|$ when we rotate the field within the honeycomb plane ($\phi$ is the field angle from the Ru-Ru bond direction). Therefore, the Majorana excitation spectrum has a finite gap for the field parallel to $\bm{a}$ or $-\bm{a}$ but it changes to gapless for $\bm{H}\parallel\bm{b}$, as shown in Fig.\,\ref{fig:schematic}\FigCap{b}. The Chern number $Ch(\bm{H})$ is given by the sign of $h_xh_yh_z$, and it is $+1$ ($-1$) for $\bm{H}\parallel -\bm{a}$ ($\bm{H}\parallel\bm{a}$) and 0 for $\bm{H}\parallel\bm{b}$ as $\bm{b}\parallel[\bar{1}10]$ is perpendicular to the $S^z$-axis (see Fig.\,\ref{fig:schematic}\FigCap{a}). Thus, when the field is rotated within the plane, the sign change of $\kappa_{xy}$ occurs in the $\bm{b}$ direction ($\phi=0^\circ$), which should be accompanied by the gap closure, as manifested in the $h_xh_yh_z$ factor in the Chern number and the mass function. The closing of the gap for $\bm{H}\parallel\bm{b}$ is also a natural consequence of the $C_2$ rotational symmetry around the bond (armchair) axis (see Supplementary Material), when we focus on the irreducible representation for the idealized honeycomb layer with the point group $D_{3d}$ ($\bar{3}1m$)~\cite{Kurumaji2023}. In stark contrast, in the topological magnon case, the sign change of the Chern number of the lowest band is caused by the topological band crossing at a finite energy, which is protected by the rule that the sum of the Chern integers for all positive magnon bands is zero~\cite{Matsumoto2014}. In this case, the low-energy excitations are always gapped in the spin disordered phase (Fig.\,\ref{fig:schematic}\FigCap{c}). Thus, the topological transition between different Chern numbers, which can be induced by field-angle rotations, accompanies a complete closure of low-energy excitation gap only for the Majorana fermion case~\cite{Hwang2022,Koyama2021}.

From the above general arguments of topological bands, the key difference that can distinguish the two origin is whether the bulk low-energy excitations are gapless or gapped when the Chern number and the thermal Hall effect change sign in the bond direction ($\bm{H}\parallel \bm{b}$). To test this experimentally, we combine low-temperature specific heat measurements for the bulk excitations and thermal Hall conductivity measurements for the edge currents in magnetic fields rotating within the honeycomb plane near the $\bm{b}$ direction. It should be emphasized that these thermal measurements are powerful and more direct probes of Majorana quasiparticles in the sense that they do not require the recombination of the fractionalized spins in the Kitaev material, unlike other spin excitation probes.

\begin{figure}[t]
        \centering
        \includegraphics[width=\linewidth]{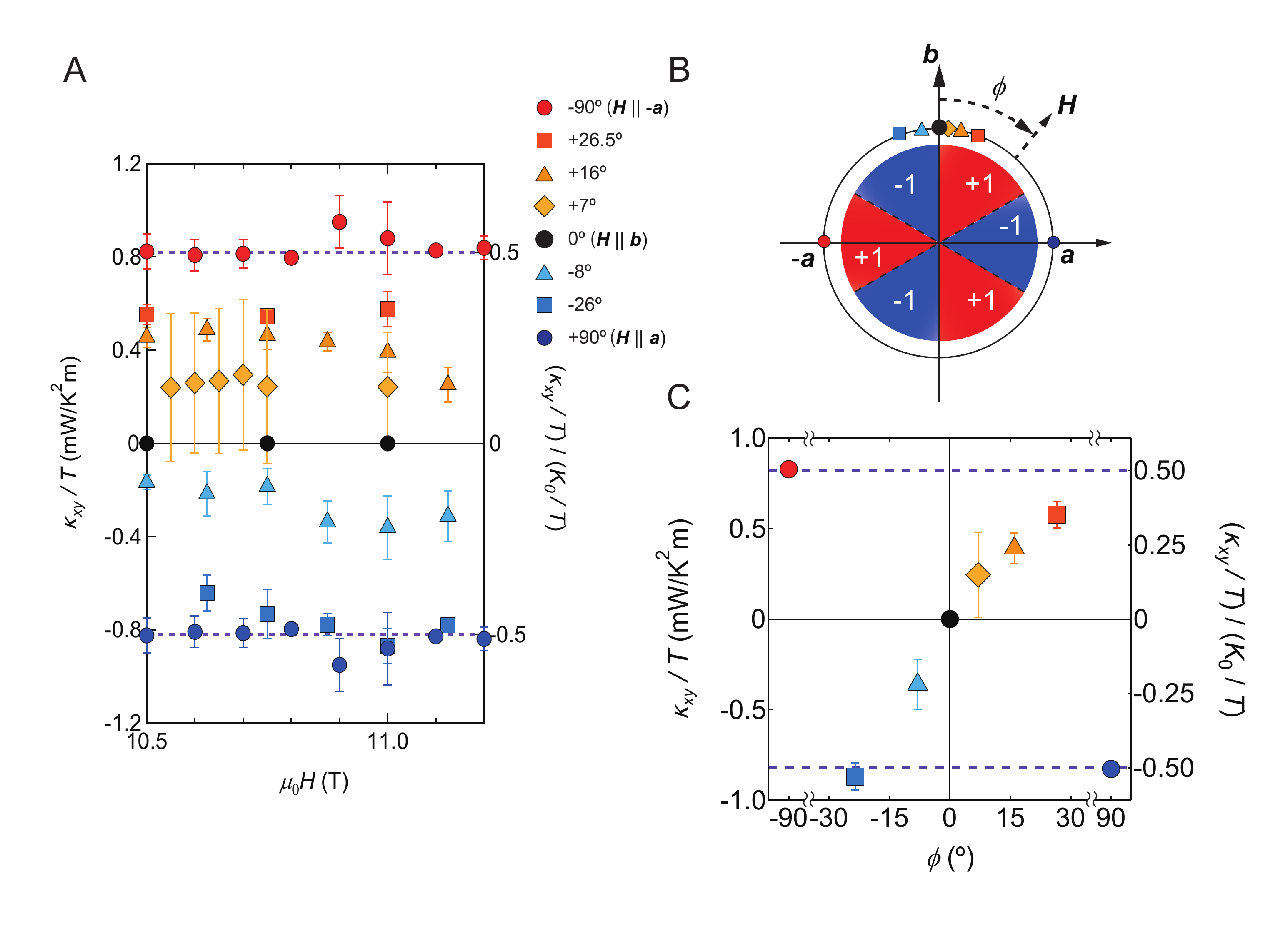}
        \caption{\label{fig:Hall}
        {\bf In-plane thermal Hall effect in the high-field quantum disordered phase of $\alpha$-RuCl$_3$.}
		({\bf \FigCap{a}}) Field dependence of $\kappa_{xy}/T$ at a temperature $T = 4.8$\,K in the in-plane magnetic fields with different angles $\phi$ from the $\bm{b}$ axis. In the right axis, thermal Hall conductance is shown in units of the quantum thermal conductance $K_0 = (\pi^2/3)(k_{\rm B}^2/h)T$. Dashed lines indicate the half-integer quantization. ({\bf \FigCap{b}}) In-plane field-angle dependence of the Chern number $Ch$. The same symbols are used as in {\bf \FigCap{a}} and {\bf \FigCap{c}} for the fixed field angles $\phi$ from the $\bm{b}$ axis. ({\bf \FigCap{c}}) Field-angle dependence of $\kappa_{xy}/T$ at a fixed field of 11\,T at $T = 4.8$\,K.} 
    \end{figure}

Figure\,\ref{fig:Hall}\FigCap{a} shows the field dependence of thermal Hall conductivity at several field angles $\phi$ defined as shown in Fig.\,\ref{fig:Hall}\FigCap{b}. In this single crystal, the half-integer quantization is seen in this field range of quantum disordered phase with a positive (negative) sign for $\bm{H}\parallel -\bm{a}$ ($\bm{H}\parallel\bm{a}$). For $\phi=-26^\circ$, $\kappa_{xy}/T\approx -0.5K_0/T$ shows a near quantization with an opposite sign from that for $\phi=-90^\circ$ ($\bm{H}\parallel -\bm{a}$), which is consistent with the six-fold sign-reversals within the plane expected for the KSL (Fig.\,\ref{fig:Hall}\FigCap{b}). When the field angle $\phi$ is further rotated from negative to positive through $\phi=0^\circ$ ($\bm{H}\parallel\bm{b}$), the sign of $\kappa_{xy}/T$ changes, which can be more clearly seen in Fig.\,\ref{fig:Hall}\FigCap{c} where the angular dependence of $\kappa_{xy}/T$ is shown at a fixed field of 11\,T. These results indicate that the topological Chern number $Ch$, which determines the direction of the chiral edge current, changes its sign in the bond direction when the field is rotated within the honeycomb plane. 

Having looked at the evidence from thermal transport measurements for the topological transition induced by field-angle rotation, we now turn to the low-energy excitation spectra in the bulk for $\bm{H}\parallel\bm{b}$. We have newly developed a high-resolution, field-angle resolved specific heat measurement system based on the long-relaxation technique~\cite{Tanaka2022}, which can cover very low temperatures down to $\sim0.2$\,K that corresponds to $\sim 1/300$ of the Kitaev interaction by utilizing a dilution fridge equipped with a two-axis piezo-based rotator (see Supplementary Material). In Fig.\,\ref{fig:Hdep}\FigCap{a}, we show the raw data of low-temperature specific heat $C$ without subtracting the phonon contribution, which is known to give a $C\propto T^3$ temperature dependence in this material~\cite{Widmann2019,Tanaka2022}. Thus, in this plot of $C/T^2$ vs.\ $T$, the phonon contribution should be described by a linear slope with no intercept in the zero-temperature limit. The temperature dependence of $C/T^2$ for $\bm{H}\parallel\bm{b}$ shows a $T$-linear behavior, but with clear residual values for $T\to0$\,K in a wide field range of the spin disordered state. This indicates the presence of non-phononic low-lying excitations that are described by $C/T =\alpha T$, where $\alpha$ is a field-dependent coefficient. This is fully consistent with the gapless excitations of a Dirac cone-like dispersion ($E=v|\bm{k}|$) in a two-dimensional (2D) system, in which the coefficient $\alpha$ is proportional to the inverse of the square of the velocity $v$. This observation of gapless excitations for $\bm{H}\parallel\bm{b}$ contradicts the expected existence of a gap in the topological magnon case (Fig.\,\ref{fig:schematic}\FigCap{c}), and provides compelling evidence for the fermionic excitations in this system. We note that the low-energy magnon gap may be closed when the system is just at a quantum critical field that separates two different magnetic phases, where 2D gapless magnons would give $T^2$ dependence of $C/T$. However, we have resolved the gapless behavior in an extended field range of the disordered phase above the antiferromagnetic critical field ($\sim 8$\,T), which indicates that the gap closure is not an accidental phenomenon but an intrinsic feature for the bond-direction fields as expected for the Majorana excitations in the KSL. The field dependence of $\alpha$ in the disordered phase shows a decreasing trend, which corresponds to the quasi-linear field dependence of Dirac velocity $v$ of the itinerant Majorana fermions (Fig.\,\ref{fig:Hdep}\FigCap{b}). This appears to be related to the effective enhancement of the Kitaev interactions with increasing field~\cite{Nasu2018}, although the quantitative understanding of $v(H)$ requires further studies. 

\begin{figure}[t]
		 \centering
		 \includegraphics[width=\linewidth]{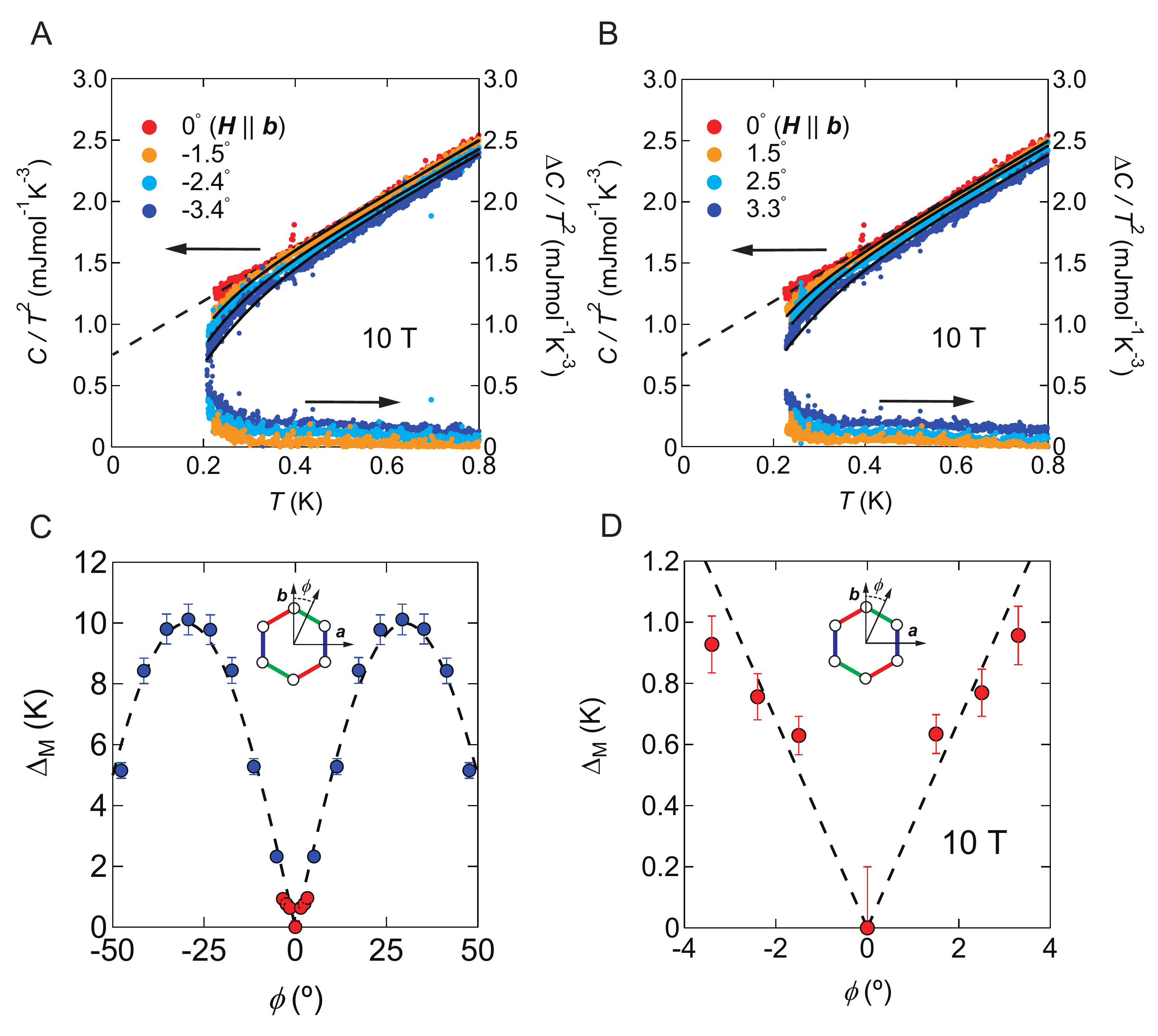}
		 \caption{\label{fig:angle}
		 {\bf Rapid opening of the Majorana gap by tilting fields from the bond direction. } 
		 ({\bf \FigCap{a}}), ({\bf \FigCap{b}}) Temperature dependence of $C/T^2$ at 10\,T for negative ({\bf \FigCap{a}}) and positive ({\bf \FigCap{b}}) tilt angles $\phi$ from the bond direction ($\bm{b}$ axis). Dashed lines represent the linear fitting for $\phi=0^\circ$ and solid lines are fitting curves with the excitation gap $\Delta_{\rm M}$ for finite angles. The data for the change in $C/T^2$ from the $\phi=0^\circ$, $\Delta C/T^2= C(\phi)/T^2- C(\phi=0^\circ)/T^2$, are also shown (right axis). ({\bf \FigCap{c}}) Angle dependence of Majorana gap $\Delta_{\rm M}$ at 10 T. The data at low angles (red circles) are obtained in this study, which are compared with the high-angle data (blue circles) reported in the previous study in the temperature range down to 0.7\,K~\cite{Tanaka2022}. $\Delta_{\rm M}(\phi)$ shows a $|\sin 3\phi|$ angle dependence (dashed lines). ({\bf \FigCap{d}}) The same data in the low-angle region. Inset shows the definition of $\phi$. 
		 }
	 \end{figure}

Further evidence for the correctness of the Majorana-fermion description can be obtained in the angular dependence of low-temperature specific heat near the bond direction. When the field is slightly inclined from the $\bm{b}$ axis, we find a deviation from the $T$-linear behavior of low-temperature $C/T^2$. Remarkably, such a deviation is found in a similar fashion for both $\phi<0^\circ$ (Fig.\,\ref{fig:angle}\FigCap{a}) and $\phi>0^\circ$ (Fig.\,\ref{fig:angle}\FigCap{b}). This can be easily understood by the fact that the observed deviations at low temperatures come from the opening of the excitation gap $\Delta_{\rm M}(\phi)\propto |\sin 3\phi|$, which is symmetric with respect to the $\bm{b}$ axis ($\phi=0^\circ$). To analyze the data quantitatively, we use the 2D model for energy dispersion $E=\sqrt{v^2|\bm{k}|^2+\Delta_{\rm M}(\phi)^2}$ for the non-phononic contribution. Then the low temperature specific heat is given by 
\begin{equation}
    C/T =\alpha T\left(1-\frac{\mathcal{G}(\Delta_{\rm M}/T)}{\mathcal{G}(\infty)}\right) +\beta T^2, \quad     
    \mathcal{G}(y)\equiv  \int_0^y \frac{{\rm d}x}{2\pi} \frac{x^3{\rm e}^x}{({\rm e}^x+1)^2}.
\label{fit} 
\end{equation}
Here, we ignore the contribution of the $Z_2$ fluxes whose gap is much larger than the measurement temperature scale, and the $\beta T^2$ term represents the phonon contribution~\cite{Tanaka2022}. The fitting gives a good agreement with the data as shown by the solid lines in Fig.\,\ref{fig:angle}\FigCap{a} and \ref{fig:angle}\FigCap{b}, with the angle-dependent gap values as demonstrated in Fig.\,\ref{fig:angle}\FigCap{c} and \ref{fig:angle}\FigCap{d}. These results obtained by the low-temperature specific heat measurements fill an important missing range in the previous higher-temperature measurements~\cite{Tanaka2022}, providing thermodynamic evidence of the distinct angle-induced gap closure. 
The angle dependence of the gap can be consistently described by the expectations for the Majorana gap $\Delta_{\rm M}(\phi)=\Delta_0|\sin 3\phi|$ with the amplitude of $\Delta_0\approx 10$\,K (Fig.\,\ref{fig:angle}\FigCap{c}). 

The obtained results of thermal Hall conductivity and specific heat, namely the sign change of the $Ch=\pm1$ chiral edge mode with the gap closure for $\bm{H}\parallel\bm{b}$ ($\phi=0^\circ$), is a signature of the topological phase transition in a fermionic system outlined in Fig.\,\ref{fig:schematic}\FigCap{b}, excluding the bosonic scenario in Fig.\,\ref{fig:schematic}\FigCap{c}. It should also be emphasized that this characteristic angle-dependent low-energy gap is not expected for the chiral phonon Hall effect. 
The combination of these measurements also demonstrates the bulk-edge correspondence of topological properties of Majorana fermion excitations; when the bulk excitations are gapped (gapless), the edge currents appear (disappear). Our results evidencing the fermionic case imply that the quantization of the thermal Hall conductivity must occur at low enough temperatures compared with the Majorana gap, if the phonon decoupling and domain issues~\cite{Vinkler2018,Ye2018,Kurumaji2023} are not at play. In our sample, the half-integer quantization plateau is indeed observed for $\bm{H}\parallel\pm\bm{a}$ ($\phi=\pm90^\circ$) at a temperature of 4.8\,K (Fig.\,\ref{fig:Hall}\FigCap{a}), which is considered high enough to neglect the phonon decoupling effect~\cite{Vinkler2018}. For small angles ($|\phi|\lesssim 10^\circ$), the Majorana gap is smaller than the measurement temperature scale for $\kappa_{xy}$ ($\Delta_{\rm M}(\phi)\lesssim 4.8$\,K) as shown in Fig.\,\ref{fig:angle}\FigCap{c}, which is likely responsible for the suppressed $\kappa_{xy}/T$ values from the half-integer quantization (Figs.\,\ref{fig:Hall}\FigCap{a} and \ref{fig:Hall}\FigCap{c}). Even without having the quantization phenomena, however, the remarkable coincidence between the bulk gap closure and the sign-changing in-plane thermal Hall effect demonstrates that the topological transition of fermionic quasiparticles is induced by field angle rotation. From these results we conclude that the excitations responsible for the planar thermal Hall effect in $\alpha$-RuCl$_3$ are itinerant Majorana fermions in the KSL.


\begin{acknowledgments}
	We thank S. Fujimoto, K. Hwang, T. Kurumaji, Y. Motome, S. Murakami, N.\ P. Ong, H. Takagi, M. Takahashi, M. Udagawa, and M.\ G. Yamada for fruitful discussions. This work was supported by CREST (JPMJCR19T5) from Japan Science and Technology (JST), Grants-in-Aid for Scientific Research (KAKENHI) (Nos.\ JP22H00105, JP22K18683, JP21H01793, JP19H00649, and JP18H05227), and Grant-in-Aid for Scientific Research on Innovative Areas “Quantum Liquid Crystals” (No.\ JP19H05824) from Japan Society for the Promotion of Science. E.-G. M. was supported by the National Research Foundation of Korea under grants NRF-2019M3E4A1080411, NRF-2020R1A4A3079707, and NRF-2021R1A2C4001847.
\end{acknowledgments}

%
    

\newpage 

	\begin{figure}[t]
		 \centering
		 \includegraphics[width=\linewidth]{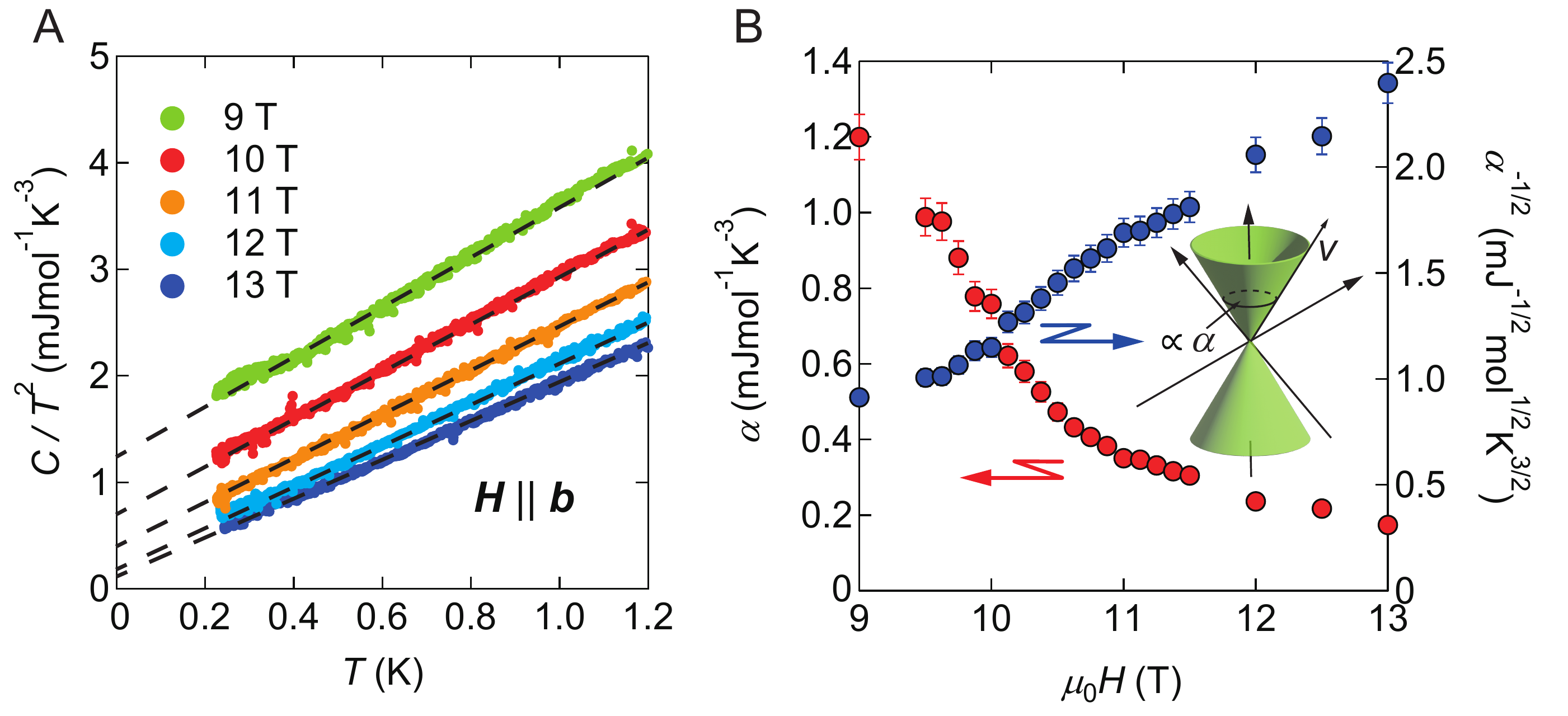}
		 \caption{\label{fig:Hdep}
         {\bf Specific-heat evidence of gapless excitations in the high-field phase of $\alpha$-RuCl$_3$ for the bond-direction fields.}
		 ({\bf \FigCap{a}}) Raw data of the temperature dependence of $C/T^2$ for $\bm{H}\parallel\bm{b}$ at several fields below 1.2\,K. The linear fitting curves (dashed lines) resolve the residual $T$-linear term of $C/T$ $\displaystyle{(\lim_{T\to0}C/T=\alpha T})$. ({\bf \FigCap{b}}) Field dependence of residual term $\alpha$ of $C/T^2$ (red circles). The right axis shows $\alpha^{-1/2}$ (blue circles), which is proportional to the Fermi velocity $v$ in the Dirac dispersion of gapless Majorana excitations (inset).} 
	 \end{figure}

\clearpage
\section*{Supplementary Materials}

\setcounter{figure}{0}
\renewcommand{\tablename}{Table S\!\!}
\renewcommand{\figurename}{Fig. S\!\!}
\renewcommand{\theequation}{S\arabic{equation}}

\subsection*{Materials and Methods}

\subsubsection{Single crystals}
High-quality single crystals of $\alpha$-RuCl$_3$ were grown by the vertical Bridgman method~\cite{Kubota2015}. We used the same sample as in Ref.~\cite{Tanaka2022}, in which the magnetic transition is seen only at 7\,K and no discernible anomaly is found at around 14\,K associated with the stacking faults. The orientation of the crystal axes is determined by X-ray diffraction measurements at room temperature. 

\subsubsection{Specific heat measurements}
The heat capacity was measured by the long-relaxation method~\cite{Tanaka2022}. In this method, we used a home-made cell consisting of a bare chip Cernox thermometer (CX1010) and silver-coated fiberglass supports (Fig.\,S1A). 
The thermometer also serves as a sample holder and a heater, which minimizes the background addenda contribution to obtain accurate temperature dependence of heat capacity even for small single crystal samples. 
The lateral size of the $\alpha$-RuCl$_3$ sample used is $\simeq 1.1\times1.3$\,mm$^2$, and the weight is $\simeq 0.68$\,mg. 
A typical relaxation curve with the sample is shown in Fig.\,S1B, where the temperature of the bare chip with and without the sample is relaxed up and down by changing the power applied to the bare chip thermometer.
A single relaxation process acquires about several hundred data points, which are converted to continuous data points of the heat capacity $C(T)$ in this temperature range through
\begin{eqnarray}
C(T)=\frac{P^{\uparrow}(T)-P^{\downarrow}(T)}{\frac{\mathrm{d}T^{\uparrow}}{\mathrm{d}t}(T)-\frac{\mathrm{d}T^{\downarrow}}{\mathrm{d}t}(T)}.
\end{eqnarray}
Here $P^{\uparrow}$ ($P^{\downarrow}$) and $\frac{\mathrm{d}T^{\uparrow}}{\mathrm{d}t}$ ($\frac{\mathrm{d}T^{\downarrow}}{\mathrm{d}t}$) is the power applied to the bare chip thermometer and the time derivative of temperature when the temperature is increased (decreased).
To achieve a high resolution, relaxation is repeated and integrated tens to thousands of times. 
Figure~S1B shows the averaged relaxation curve obtained by repeating the relaxation 1000 times.
We obtained the heat capacity data up to higher temperatures by  performing this sequence with many different based temperatures. 
The heat capacity of addenda, consisting of the chip and grease, is measured before the sample is mounted and is subtracted.
The heat capacity of the addenda is 10-100 times smaller than that of the sample in the measured temperature and field range.
\par
The heat capacity measurements were performed in a Bluefors cryogen-free dilution refrigerator equipped with a superconducting magnet enabling $C(T)$ measurements at very low temperatures down to $\sim200$\,mK and in magnetic field up to 13\,T.
The home-made cell was mounted on a two-axis rotator (attocube atto3DR), permitting to vary the angle three-dimensionally between the applied magnetic field and the sample.
The in-plane angle $\phi$ (Fig.\,S2A) was checked by two Hall sensors orthogonally placed on the rotator and the out-of-plane angle $\theta$ was determined by the resistive sensor originally installed in the atto3DR system with a sensor resolution of about 6\,mdeg.
\par
We align the direction of the magnetic field to the $\bm{b}$ axis by using the angle dependence of specific heat $C/T$. 
Figure~S2B shows the in-plane and out-of-plane angle dependence of specific heat $C/T$ at fixed field magnitudes at 0.4~K around the $\bm{b}$ axis.
As shown by the red circles, $C/T$ shows strong $\phi$ dependence when magnetic field is rotated.
$C/T$ exhibits the six-fold oscillation with respect to the angle $\phi$ and has the maxima along the $\bm{b}$ axis, corresponding to the anisotropy of the Majorana gap $\Delta_M\propto{|\sin3\phi|}$~\cite{Tanaka2022}.
Therefore, the angle at which $C/T$ reaches the maximum value is considered to be the angle at which the magnetic field is parallel to the $\bm{b}$ axis.
After $\phi$ is adjusted to 0$^\circ$ this way, we measured $C/T$ as a function of $\theta$, which shows almost no $\theta$ dependence between $-5^{\circ}$ and $5^{\circ}$ (blue circles in Fig.\,S2B).
This implies that the magnetic torque can bend the thin fiberglass suspending the sample stage, so that the sample plane is always locked to the field direction within this $\theta$ range. Indeed, the $\theta$-independent $C/T$ value stays the maximum value of $C/T(\phi)$, which indicates that the field direction is precisely aligned along the $\bm{b}$ direction.

\subsubsection{Thermal transport measurements}
For thermal conductivity measurements, Crystal \#3 in Ref.~\cite{yokoi2021half}, in which the half-integer quantized thermal Hall effect is observed, was used. The direction of the crystal axes was determined by Laue X-ray back reflection measurements. The size of the crystal is 2150 $\mathrm{\mu m}$ (length) $\times$ (1000 $\pm$ 10) $\mathrm{\mu m}$ (width) $\times$ (20 $\pm$ 0.3) $\mathrm{\mu m}$ (thickness).

The thermal conductivity cell is shown in Fig.\,S3. An insulating LiF plate fixed to a large Cu block was used as a heat bath for the crystal. One end of the crystal was glued to the LiF plate, which serves as the heat bath, with a non-metallic grease. A 1-k$\Omega$ resistance chip was used as a heater. Three Cernox thermometers (CX1070), which were carefully calibrated in magnetic fields, were used to measure the thermal gradient. Thermal current $\bm{j}$ was applied along the crystallographic $\bm{a}$ axis. In thermal Hall measurements, three thermometers were attached to the crystal to simultaneously measure the longitudinal ($\nabla T_x$) and the transverse ($\nabla T_y$) thermal gradients. The direction of the in-plane magnetic field was rotated by rotating the thermal conductivity cell in a superconducting magnet. Since the thermal Hall signal is an odd function of the magnetic field, we measured the thermal gradient for positive and negatve $H$,  and anti-symmetric components of the measured transverse temperature gradient were numerically calculated and used to obtain $\kappa_{xy}(H)$.



\subsection*{Discussion about the symmetry properties}
The concomitant energy-gap closing ($\Delta_M \propto | M(\bm{H}) | = 0 $) and the zero thermal Hall conductivity ($\kappa_{xy}=0$) under magnetic fields along the bond directions are related to the symmetry properties.
In the idealized monolayer structure of $\alpha$-RuCl$_3$~\cite{Kurumaji2023} as in the original Kitaev model~\cite{Kitaev2006}, the group symmetry is $D_{3d}$ (or $\bar{3}1m$) symmetry consisting of three-fold rotational symmetry with respect to the $c$-axis ($C_3$), three two-fold rotations with respect to the bond directions ($C_2$), and the inversion symmetry ($P$). 
Since the magnetic field is even under $P$, it is sufficient to focus on $D_3$ symmetry whose character table and basis functions are summarized in Tables SI and SII. 

\begin{table}[b]
\caption{The character table for the $\bm{D_3}$ group.}
    \begin{tabular}{>{\centering}p{1cm}|>{\centering}p{2cm}|>{\centering}p{2cm}|>{\centering\arraybackslash}p{2cm}}
\noalign{\smallskip}\noalign{\smallskip}\hline\hline
& 1 & $3C_2$ & $2C_3$ \\
\hline
$\mathcal{A}_1$ & 1 & 1 & 1 \\
\hline
$\mathcal{A}_{2}$& $1$ & $-1$ &$1$\\
\hline
$\mathcal{E}$  & $2$& $0$ & $-1$  \\
\hline
\hline
\end{tabular}
\end{table}

\begin{table}[t]
\caption{The basis functions with magnetic field, $\bm{H} = (h_a, h_b, h_c)$ in the crystallographic coordinate. }
\begin{tabular}{>{\centering}p{1cm}|>{\centering}p{2cm}|>{\centering\arraybackslash}p{9cm}}
\noalign{\smallskip}\noalign{\smallskip}\hline\hline
& \text{order 1} & \text{order 3}  \\
\hline
$\mathcal{A}_1$ & - & $h_b(h_b^2-3h_a^2)$  \\
\hline
$\mathcal{A}_{2}$& $h_c$ &$h_a(h_a^2-3h_b^2), h_c^3, h_c(h_a^2+h_b^2)$ \\
\hline
$\mathcal{E}$  &  $[h_a,h_b]$& $h_c^2[h_a,h_b], (h_a^2+h_b^2)[h_a,h_b], h_c^2[h_a^2-h_b^2,h_ah_b]$    \\
\hline
\hline
\end{tabular}
\end{table}

The thermal conductivity $\kappa_{xy}$ and the mass function $M(\bm{H})$ are in the $\mathcal{A}_{2}$ representation, and along the bond directions of magnetic fields (for example, $h_a=h_c=0$), they become zero because they are odd under the $C_2$ symmetries. Note that the mass function for the pure Kitaev model with magnetic fields is~\cite{Kitaev2006}, 
\begin{eqnarray}
M(\bm{H}) \propto  -2h_a(h_a^2-3h_b^2) +2h_c^3-3\sqrt{2}h_c(h_a^2+h_b^2)= 6\sqrt{3}h_x h_y h_z \in \mathcal{A}_2.
\end{eqnarray}
%
%
%

We note that three candidates for the crystal structure of bulk $\alpha$-RuCl$_3$ have been suggested at low temperatures, $R\bar{3}$~\cite{Park,Nagai2020}, $P3_112$, and $C2/m$~\cite{bruin2022origin}, while the symmetry of the high-temperature phase of the Bridgeman crystals is known to be $C2/m$~\cite{bruin2022origin}. The three structures have different layer stacking arrangements and their symmetries are lower than the idealized monolayer symmetry $D_{3d}$. 
For $P3_112$, one of the subgroup is $D_{3}$, and thus $\kappa_{xy} = \Delta_M =0$ for the bond directions of magnetic fields. For $R\bar{3}$, the two representations $\mathcal{A}_1$ and $\mathcal{A}_2$ are mixed, yet the existence of the magnetic field directions for the gap closing is guaranteed, shifted from the bond directions. The shift is expected to be small when the interlayer coupling is weak, which is the case for $\alpha$-RuCl$_3$ with weak van der Waals stacking of the honeycomb layers.
We further note that if the point group symmetry is lowered to $C2/m$, only one of the bond directions (along the $\bm{b}$ axis) gives the gapless condition.


\begin{figure}[t]
    \centering
    \includegraphics[width=0.8\linewidth,pagebox=artbox]{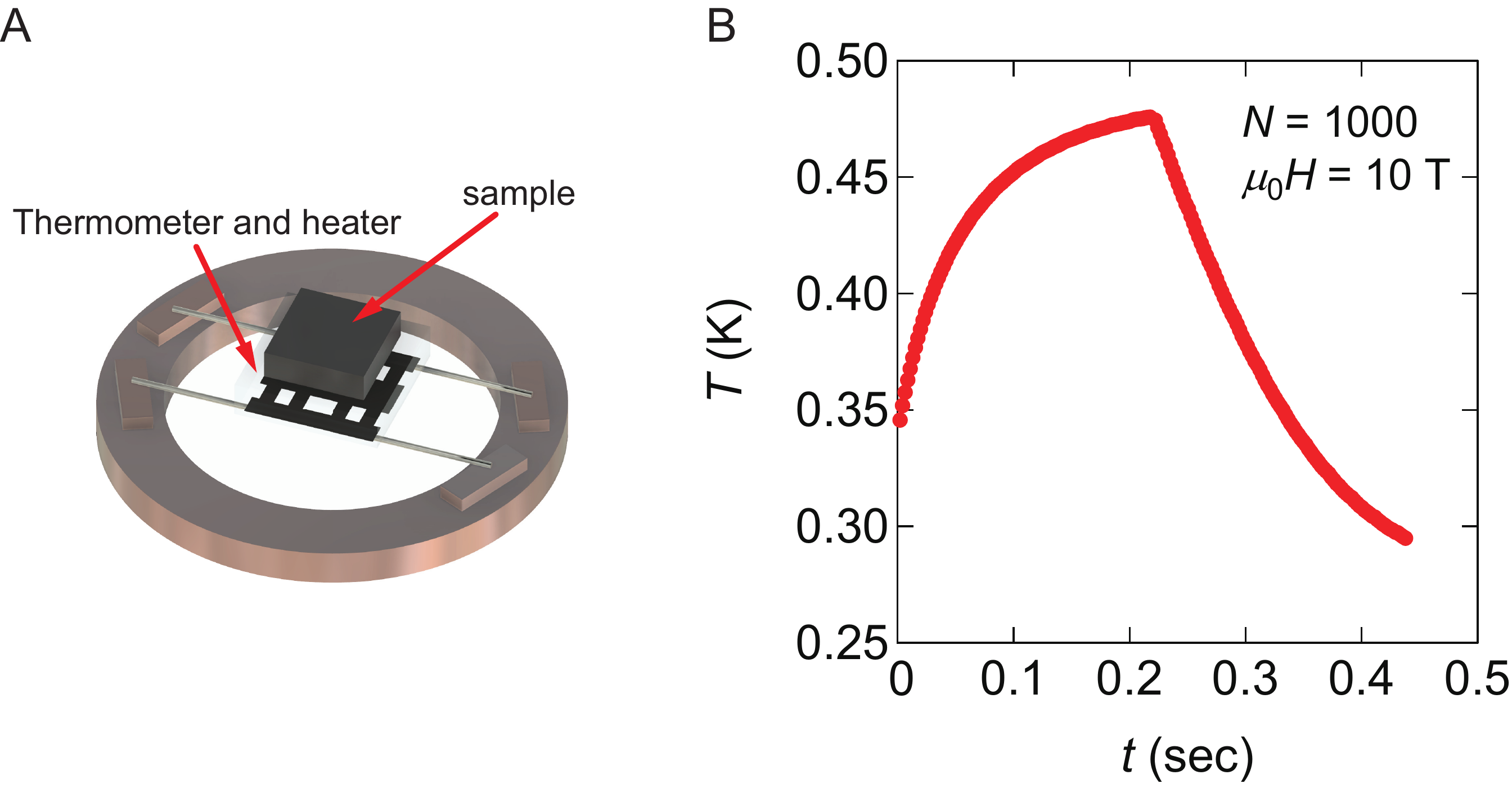}
    \caption{{\bf Specific heat measurement setup.} 
    ({\bf A}) Schematic of experimental set up for the specific heat measurements. ({\bf B}) Typical averaged relaxation curve with the $\alpha$-RuCl$_3$ single crystal at very low temperatures obtained by repeating the heating and cooling processes 1000 times.
    }
    \label{Cfitting}   
\end{figure}

\begin{figure}[b]
    \centering
    \includegraphics[width=0.8\linewidth,pagebox=artbox]{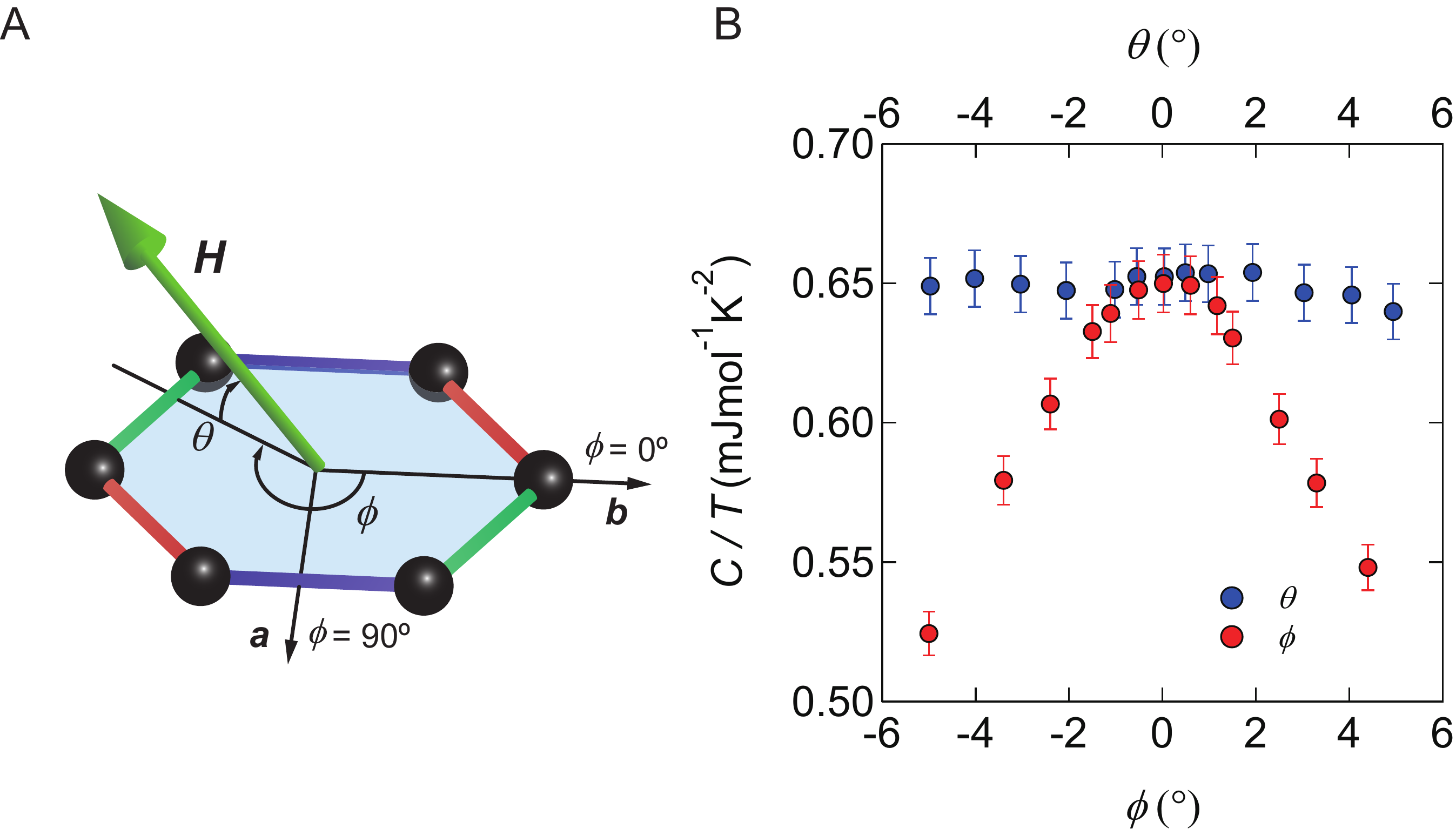}
    \caption{{\bf Field angle alignment by using a two-axis rotator.}
    ({\bf A}) The direction of magnetic field $\bm{H}$ (green arrow) is controlled by a two-axis rotator. The definitions of polar angle $\theta$ and azimuthal angle $\phi$ are shown.     
    ({\bf B}) Angle dependence of $C/T$ at 10\,T at 0.40\,K. From the $\phi$-scan measurements, $\phi=0^\circ$ is determined as the angle at which $C/T$ shows the maximum value, and the $\theta$-scan measurements are performed for $\phi=0^\circ$.  
    }
    \label{Ccalc}   
\end{figure}

\begin{figure}[t]
    \centering
    \includegraphics[width=0.38\linewidth,pagebox=artbox]{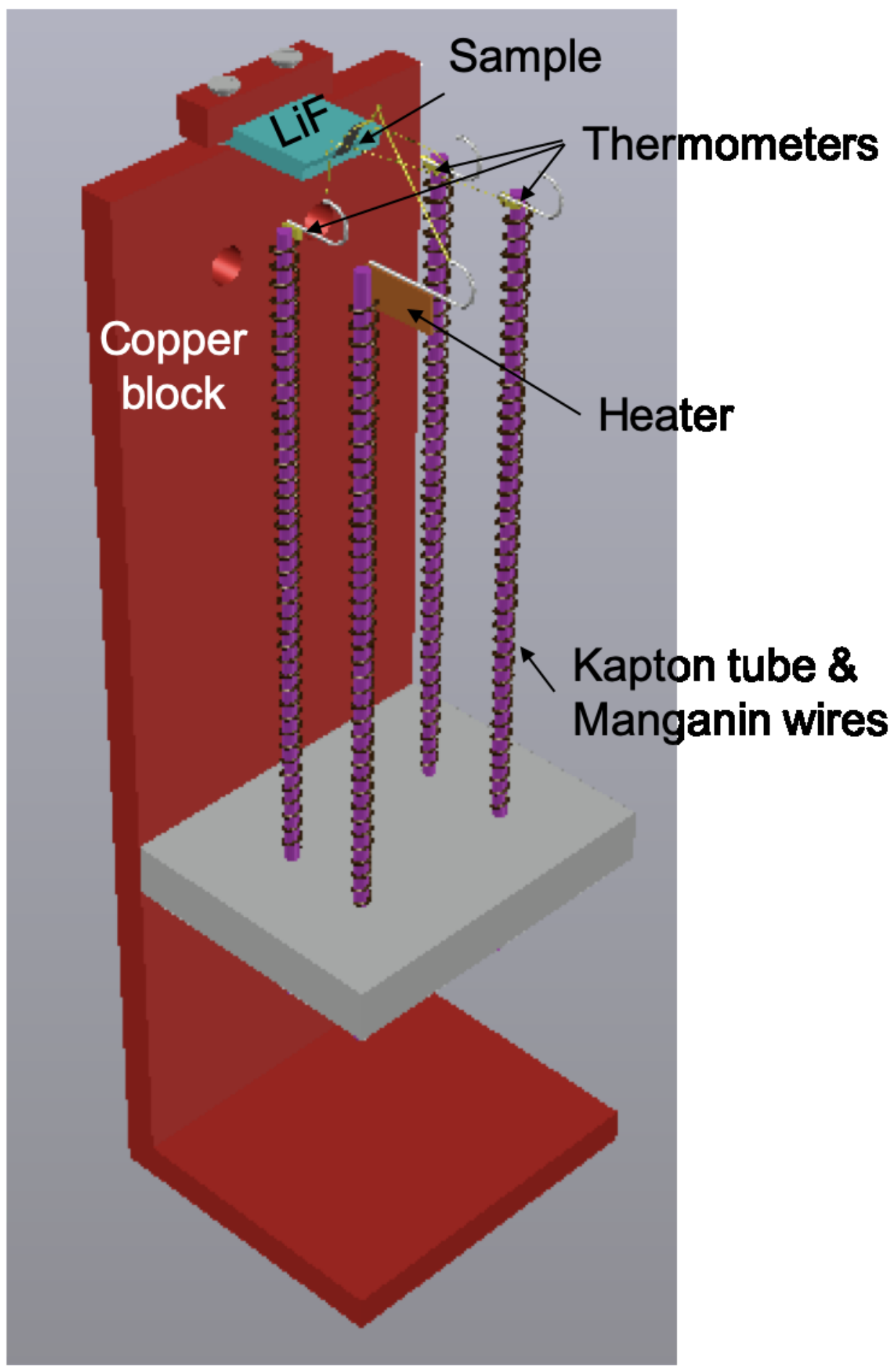}
    \caption{{\bf Schematic of experimental set up for thermal conductivity measurements.} 
    }
    \label{kappa}   
\end{figure}

\end{document}